\documentstyle[prl,aps,multicol,amssymb,epsfig]{revtex}
\begin{document}
\title{Electron spin relaxation in semiconductors and semiconductor structures}
\author{Yu.G.Semenov}
\address{Groupe d'Etude des Semi-Conducteurs, Universit\'{e}
Montpellier 2,\\
Place Eug\`{e}ne Bataillon, 34095 Montpellier, France}
\date{\today }
\maketitle

\begin{abstract}

We suggest an approach to the problem of free electron spin evolution in a
semiconductor with arbitrary anisotropy or quantum structure in a magnetic
field. The developed approach utilizes quantum kinetic equations for average
spin components. These equations represent the relaxation in terms of
correlation functions for fluctuating effective fields responsible for spin
relaxation. In a particular case when autocorrelation functions are
dominant, the kinetic equations reduce to the Bloch equations. The developed
formalism is applied to the problem of electron spin relaxation due to
exchange scattering in a semimagnetic quantum well (QW) as well as to the
spin relaxation in a QW due to Dyakonov-Perel mechanism. The results permit
to separate the longitudinal $T_1$ and transversal $T_2$ relaxations times
in a strong enough magnetic field and to trace the cases of undistinguished
parameters $T_1$ and $T_2$ in zero and small magnetic fields. Some new
predictions of the developed theory are discussed.
\end{abstract}


\begin{multicols}{2}

\narrowtext

\section{Introduction}

The achievements in the investigations of ferromagnetism (FM) of
diluted magnetic semiconductors (DMS) \cite{OhnoReview} and in
spin injection technology\cite{OhnoInject} with the possibility to
fabricate the spin-controlled devices have sparked renewed
interest to studies of spin relaxation of electrons, holes and
excitons (see \cite{Vina} and references therein). During the
recent years, the optical picosecond technique acquire ability to
observe the electron and hole spin kinetics in semiconductor
quantum structures over wide range of temperatures and magnetic
fields \cite {Crooker1},\cite{Crooker2}. In spite of intensive
work the spin relaxation processes are still not fully understood
in these structures. One of the reasons is the deficiencies in the
present theories. If, for example, the theory of giant spin
splitting of the electron states in DMS has a good base in terms
of molecular exchange field, \cite{FK}, spin relaxation theory
has, by now, a huge gap between phenomenological and microscopic
approaches to this problem.

The phenomenological description of carrier or exciton spin relaxation in
semiconductors is commonly started with Hamiltonian $H_d$ without
dissipation terms responsible for spin relaxation. For a particle with spin $%
S$ (electron, exciton, etc.), the number of equations for average
spin values $A_I=<S_X^kS_Y^lS_Z^m>$
(subscripts denote Cartesian components; $k,l,m$ satisfy the condition $%
1\leq k+l+m\leq 2S$ ; symbol $I$ marks all indexes in the right
hand part of $A_I$) is finite and is determined by the equations
of motion ( $\hbar =1$)

\begin{equation}
\frac{dA_I}{dt}=-i\left[ A_I,H_d\right] .  \label{eq1}
\end{equation}
The system (\ref{eq1}) describes the evolution of spin system without
relaxation. To take it into consideration we should include the relaxation
part $R_I$ to the Eq. (\ref{eq1}). Generally speaking, the relaxation of
each $A_I$ to thermal equilibrium value $A_J^0$ might depend on full set of
spin averages. Therefore relaxation part reads

\begin{equation}
R_I=-\sum_J\frac{(A_J-A_J^0)}{\tau _{I,J}}.  \label{eq2}
\end{equation}
Sometimes, the number of relaxation parameters $\tau _{I,J}^{-1}$ can be
reduced due to the presence of specific physical mechanisms (see, for
example,\cite{DyakBrus}). Nevertheless, it should be mentioned that a
general procedure (for estimation of magnitudes of all relaxation
parameters) that allows to reduce the numbers of relaxation terms (\ref{eq2}%
) is absent.

Other problem arises when one tries to relate the parameters $\tau
_{I,J}^{-1}$ with the spin-flip rate $W$ (this is the aim of most
microscopic calculations based on different known relaxation mechanisms). To
clarify this problem we consider Bloch equations for average spin components
$S_\mu $ in a magnetic field directed along $OZ$ axis
\begin{equation}
\frac d{dt}\left(
\begin{array}{c}
S_X \\
S_Y \\
S_Z
\end{array}
\right) =\omega \left(
\begin{array}{c}
-S_Y \\
S_X \\
0
\end{array}
\right) -\left(
\begin{array}{c}
S_X/T_2 \\
S_Y/T_2 \\
(S_Z-S_Z^0)/T_1
\end{array}
\right) \text{.}  \label{eq2a}
\end{equation}
Eqs (\ref{eq2a}) looks like Eq.(\ref{eq1}),(\ref{eq2}), where
$S_Z^0$ is a thermal equilibrium value of $S_Z$, $\omega $ is
Zeeman splitting in a magnetic field. Eqs (\ref{eq2a}) involve two
relaxation times, the longitudinal $T_1$ and transversal $T_2$. In
the framework of phenomenological Eqs (\ref{eq2a}) it is not
apparent how the parameters $T_1$ and $T_2$ having different
physical meaning are related to spin-flip probability $W$.
Moreover, the transformation from the case of $T_1=T_2$ at zero
magnetic field ($\omega =0)$ to the case $T_1$ $\neq $ $T_2$ at
$\omega \neq 0$ cannot be traced both with the help of Eqs
(\ref{eq2a}) and by the microscopic calculations of spin-flip rate
$W$.

The present work is an attempt to ''fill the gap'' between the
phenomenological approach to spin relaxation and microscopic calculations of
spin flip rates in the case of electron relaxation with spin $1/2$ in a
semiconductor. To do that, we derive the quantum kinetic equations for
density matrix of spin $1/2.$ The equations have been derived in terms of
correlation functions of dissipative subsystem. Despite the presence of some
symmetry of correlation parameters, the kinetic equations can be only
reduced to the form that remains, nevertheless, more complicated than Bloch
equations. In the following sections the developed theory is applied to
electron exchange scattering on the magnetic ions in a semimagnetic quantum
well (QW). Then, Dyakonov-Perel mechanism is considered with respect to
anisotropy of relaxation times in a QW and their reduction in a magnetic
field. We discuss also the possible applications of the results obtained here.

\section{Basic equation for density matrix}

Let the electron spins represent a \textit{small part} of total system so
that all other variables is related to the bath being at thermal
equilibrium. Corresponding Hamiltonian reads
\begin{equation}
H=H_S^0+H_L+H_{SL}.  \label{eq3}
\end{equation}
Here $H_S^0$, $H_L$, and $H_{SL}$ are, respectively, the Hamiltonians of
electron spin, dissipative subsystem (assumed to be at thermal equilibrium
with a bath at a temperature $T$), and their interaction. According to
projection operator method (Ref.\cite{Zwanz}) the density matrix of the
system (\ref{eq3}) is expressed in terms of main part and the rest:
\begin{equation}
\rho =f\sigma +\eta ,  \label{eq4}
\end{equation}
where
\begin{eqnarray*}
\sigma &=&{\rm Tr}_L\rho ; \\
f &=&\frac{\exp \left( -\beta H_L\right) }{{\rm Tr}_L\exp \left( -\beta H_L\right)
}; \\
\eta &=&P\rho \equiv (1-f{\rm Tr}_L)\rho .
\end{eqnarray*}

We are looking for the kinetic equation for reduced density matrix $\sigma $
with the Hamiltonian

\begin{equation}
H=H_S+H_L+V,  \label{eq5}
\end{equation}
where
\begin{eqnarray}
H_S &=&H_S^0+\left\langle H_{SL}\right\rangle ;  \label{eq5a} \\
V &=&H_{SL}-\left\langle H_{SL}\right\rangle ,  \nonumber
\end{eqnarray}
so that $\left\langle V\right\rangle =0$, where $\left\langle
...\right\rangle ={\rm Tr}_Lf...$..

It can be shown that exact form of equation for the $\sigma $ is
\begin{eqnarray*}
\frac{d\sigma }{dt}=-i{\cal H}_S\sigma -{\cal C}\left( \sigma \right)
-i{\cal D}\left( \rho \left( 0\right) \right) ;
\end{eqnarray*}
with
\begin{eqnarray}
&&{\cal C}\left( \sigma \right) ={\rm Tr}_L{\cal V}\int _0^t{\cal S}%
\left( t,t'\right) {\cal V}f\sigma \left( t'\right)
dt',  \nonumber \\
&&{\cal D}\left( \rho \left( 0\right) \right) ={\rm Tr}_L{\cal V}{\cal S}\left(
t,0\right) P\rho \left( 0\right) ,  \nonumber \\
&&{\cal S}\left( t,t'\right) =\exp \left\{
-i\int_{t'}^t\left( {\cal H}_S\left( \tau \right) +%
{\cal H}_L+P{\cal V}\right) d\tau \right\} .  \label{eq6}
\end{eqnarray}

Here the calligraphic letters mean the Liouville operators introduced over
operator space as ${\cal H}_S\sigma =\left[ H_S,\sigma \right] $, etc.
The term ${\cal D}\left( \rho \left( 0\right) \right) $ vanishes for the
initial condition $\rho \left( 0\right) =f\sigma \left( 0\right) $.

The expansion of Eq.(\ref{eq6}) up to second order in the operator of
interaction yields the kinetic equation for $\sigma $:

\begin{equation}
\frac{d\sigma \left( t\right) }{dt}=-i{\cal H}_S\sigma
-{\rm Tr}_L\int_0^t{\cal V}\left( t,t\right) {\cal V}\left(
t,t'\right) f\sigma \left( t\right) dt',  \label{eq7}
\end{equation}

where

\begin{eqnarray}
&&V\left( t,t'\right) =\exp \left\{ -i\int\limits_{t^{\prime
}}^tH_S\left( \tau \right) d\tau \right\} \times \nonumber \\
&&\times V\left( t\right) \exp \left\{
i\int\limits_{t^{\prime }}^tH_S\left( \tau \right) d\tau \right\} ,
\label{eq8}\\
&&V\left( t\right) =\exp \left\{ iH_Lt\right\} V\exp \left\{ -iH_Lt\right\}
.  \label{eq8a}
\end{eqnarray}
We consider the reduced density matrix $\sigma (t)$ at the long enough time
scale (as compared to times of attainment of a thermal equilibrium of
dissipative subsystem). Thus, the upper limit of integration in Eq. (\ref
{eq7}) can be extended to infinity. Under this assumption we obtain Markovian
equation for $\sigma (t)$, providing that spin relaxation times $T$ are
sufficiently longer than correlation times $\tau $ of dissipative system
responsible for spin relaxation.

To rewrite Eq.(\ref{eq7}) in the matrix form, we should use some base. As
the base, here we use the eigenvectors of spin operator $S_Z$ . For $S=1/2$,
the most general form of interaction is:\cite{Pikus}

\begin{equation}
H_{SL}=\bf {S\Omega },  \label{eq9}
\end{equation}
while electron spin-Hamiltonian can be written as

\begin{equation}
H_S=\omega S_Z\text{,}  \label{eq10}
\end{equation}
$\omega $ is the spin splitting in a magnetic field. Eq. (\ref{eq10}) can be
considered also as a definition of $OZ$ axis as quantization direction.
Substitution of Eq.(\ref{eq9}) to Eq.(\ref{eq7}) (at $t\to \infty )$%
, permits to express the collision integral in terms of the different
Fourier components of correlation functions for $\Omega _\alpha $. Their
number can be reduced with respect to the spectral properties of correlation
functions and to the following identities (valid for any operators $A$ and $%
B)$:

\begin{equation}
\left\langle A\left( \tau \right) B\right\rangle _\omega =e^{-\beta \omega
}\left\langle B\left( \tau \right) A\right\rangle _{-\omega }=e^{-\beta
\omega }\left\langle BA\left( \tau \right) \right\rangle _\omega ,
\label{eq11}
\end{equation}
where the definitions of $A(t)$ and $B(t)$ are similar to Eq. (\ref{eq8a}); $%
\beta =1/T$. The definition of Fourier transformation of correlation
function is
\begin{equation}
\left\langle A\left( \tau \right) B\right\rangle _\omega =\frac 1{2\pi
}\int\limits_{-\infty }^\infty \left\langle A\left( \tau \right)
B\right\rangle e^{i\omega \tau }d\tau .  \label{eq11a}
\end{equation}

It is convenient to introduce the average values of spin components instead
of density matrix $\sigma $:
\begin{eqnarray}
\left\langle S_X\right\rangle  &\equiv &{\bf X}={\rm Tr}\sigma S_X,  \label{eq12}
\\
\left\langle S_Y\right\rangle  &\equiv &{\bf Y}={\rm Tr}\sigma S_Y,  \nonumber \\
\left\langle S_Z\right\rangle  &\equiv &{\bf Z}={\rm Tr}\sigma S_Z.  \nonumber
\end{eqnarray}
Straightforward calculations with the aid of Eqs(\ref{eq11},\ref{eq11a})
yield following system of equations for components of $\left\langle \mathbf{S%
}\right\rangle $:

\begin{equation}
\frac d{dt}\left(
\begin{array}{c}
{\bf X} \\
{\bf Y} \\
{\bf Z}
\end{array}
\right) =\omega \left(
\begin{array}{c}
-{\bf Y} \\
{\bf X} \\
0
\end{array}
\right) -\mathbf{\Gamma }\left(
\begin{array}{c}
{\bf X} \\
{\bf Y} \\
{\bf Z}-{\bf Z}_0
\end{array}
\right),  \label{eq13}
\end{equation}
where relaxation matrix ${\bf \Gamma }$ is determined via correlation
functions
%$\mathbf{X}\stackunder{t\rightarrow \infty }{\rightarrow }\mathbf{X}_0$
\begin{equation}
\gamma _{\mu \nu }\left( \omega \right) =\left\langle \Omega _\mu
\left( \tau \right) \Omega _\nu \right\rangle _\omega .
\label{eq11ab}
\end{equation}
Equation (\ref{eq13}) describes the relaxation of spin 1/2 to its
equilibrium state ${\bf X}\left( t\right) \stackrel{t\to \infty
}{\longrightarrow }{\bf X}_0$,${\bf Y}\left( t\right)
\stackrel{t\to \infty }{\longrightarrow }{\bf Y}_0$, and ${\bf
Z}\left( t\right)
\stackrel{t\to \infty }{\longrightarrow }{\bf Z}_0$, with ${\bf X}_0=%
{\bf Y}_0=0,$ and
\begin{equation}
{\bf Z}_0=-\frac 12\tanh \frac{\beta \omega }2.  \label{eq11b}
\end{equation}

The general form of relaxation matrix ${\bf \Gamma }$ is quite complex.
Below, we consider two important specific cases permitting to simplify the
components of correlation matrix $\gamma _{\mu \nu }\left( \omega \right) $ (%
\ref{eq11ab}); $\nu ,\mu =X,Y,Z$. First, we assume that the matrix
(\ref{eq11ab}) is symmetric, $\gamma _{\mu \nu }\left( \omega
\right) =\gamma _{\nu \mu }\left( \omega \right) $. Thus, the
magnetic field influence on Hamiltonian of dissipative subsystem
is disregarded (Ref. \cite{Landau}). With respect to symmetric
property of correlation function, one can obtain
\begin{equation}
\mathbf{\Gamma =}\frac \pi \hbar \left(
\begin{array}{ccc}
\gamma _{zz}^0+n\gamma _{yy} & -n\gamma _{xy} & -n\gamma _{xz} \\
-n\gamma _{xy} & \gamma _{zz}^0+n\gamma _{xx} & -n\gamma _{yz} \\
-n\gamma _{xz} & -n\gamma _{yz} & n(\gamma _{xx}+\gamma _{yy})
\end{array}
\right) \text{;}  \label{eq14}
\end{equation}
$n\equiv n(\omega )=\left( 1+e^{\beta \omega }\right) /2$, $\gamma _{\mu \nu
}\equiv \gamma _{\mu \nu }\left( \omega \right) ,\gamma _{\mu \nu }^0\equiv
\gamma _{\mu \nu }\left( 0\right) $ . Notice that Eqs(\ref{eq13}),(\ref{eq14}%
) are, generally spiking, derived for the case of anisotropic medium and
arbitrary spin splitting $\omega $.

In the case of zero spin splitting ($\omega =0$) and isotropic medium, the
relaxation matrix $\mathbf{\Gamma }$ (\ref{eq14}) becomes symmetric. In this
case, Eq.(\ref{eq14}) are similar to those obtained in the work \cite{Pikus}
in terms of fluctuating internal magnetic field $\mathbf{\Omega }$.

In the case of strong enough spin splitting, one can expect that $\gamma
_{\alpha \alpha }(\omega )<<\gamma _{zz}(0)$. If we introduce $T_1^{-1}=\pi
n\left( \gamma _{xx}(\omega )+\gamma _{yy}(\omega )\right) $ and $%
T_2^{-1}=\pi \gamma _{zz}(0)$, and omit the off-diagonal components of $%
\Gamma $, the latter inequality permits to rewrite the Eq.(\ref{eq13} ) to
the form of Bloch equations (\ref{eq2a}).

Second important case corresponds to antisymmetric coefficients
$\gamma _{\mu \nu }\left( \omega \right) =-\gamma _{\nu \mu
}\left( \omega \right) $ for $\nu \neq \mu .$ In this case we can
obtain other form of relaxation matrix:
\end{multicols}
\widetext
\noindent\rule{20.5pc}{0.1mm}\rule{0.1mm}{1.5mm}\hfill

\begin{equation}
\mathbf{\Gamma =}\frac \pi \hbar \left(
\begin{array}{ccc}
\gamma _{zz}^0+n\gamma _{yy}+ni\gamma _{xy} & 0 & ni\gamma _{yz} \\
0 & \gamma _{zz}^0+n\gamma _{xx}+ni\gamma _{xy} & ni\gamma _{zx} \\
0 & 0 & n(\gamma _{xx}+\gamma _{yy}+i\gamma _{xy})
\end{array}
\right) \text{.}  \label{eq14a}
\end{equation}

\hfill\rule[-1.5mm]{0.1mm}{1.5mm}\rule{20.5pc}{0.1mm}
\begin{multicols}{2}
\narrowtext
In the limit of zero magnetic field, the coefficients $\gamma _{\mu \nu
}\left( \omega \rightarrow 0\right) $ vanish if $\nu \neq \mu $; thus the
Eq.(\ref{eq14a}) transforms to matrix (\ref{eq14}) with zero off-diagonal
components. The spectral properties (\ref{eq11}) with respect to definition
of $n(\omega )$ permits to observe that both relaxation matrix (\ref{eq14a})
and matrix (\ref{eq14}) are even functions of the electron spin splitting $%
\omega $: $\mathbf{\Gamma }(\omega )=\mathbf{\Gamma }(-\omega )$.

The Eqs (\ref{eq13}) with definitions (\ref{eq14}),(\ref{eq14a})
are the main result of theoretical background of the work. In
derivation of kinetic equations (\ref{eq13}) we assume neither any
specific properties of heat bath with Hamiltonian $H_L$ nor any
specific interaction mechanism $H_{eL}$. We shall show how Eqs
(\ref{eq13}) can be used in the particular cases corresponding to
(\ref{eq14}) and (\ref{eq14a}). The short review of the relations
between correlation functions (\ref{eq11a}) and two-time
temperature Green's functions (with well-known calculation
technique) is presented in Appendix.

\section{Exchange scattering on magnetic ions in the 2D QW}

Consider the problem of electron spin exchange scattering on the magnetic
ions in a 2D quantum well (see Ref.\cite{Bastard1}\cite{SK}). The
carrier-ion spin-spin interaction operator has the form of following contact
interaction
\begin{equation}
H_{SL}=-\alpha \sum\limits_j\mathbf{S}^j\mathbf{s}_e\delta (\mathbf{r}-%
\mathbf{R}_j),  \label{eq19}
\end{equation}
where $\alpha $ is exchange integral, $\mathbf{S}^j$ is the magnetic ion
spin at the site $\mathbf{R}_j$; $\mathbf{s}_e$ and $\mathbf{r}$ are the
spin and coordinate of electron respectively. The sum is performed over all
magnetic impurities. The Hamiltonian of dissipative subsystem include Zeeman
energy of magnetic ions and the electron kinetic energy $\varepsilon _{%
\mathbf{k}}$ with in-plane wave vector $\mathbf{k}$. For simplicity sake, we
restrict our consideration by the ground electronic confinement state only.
So, the Hamiltonian is assumed to take the form:
\begin{equation}
H_L=\sum\limits_j\omega _0S_Z^j+\sum\limits_{\mathbf{k},\sigma }\varepsilon
_{\mathbf{k}}a_{\mathbf{k},\sigma }^{\dagger }a_{\mathbf{k},\sigma },
\label{eq20}
\end{equation}
where $\omega _0$ is the magnetic ion Zeeman splitting in a magnetic field
directed along $OZ$ axis; $a_{\mathbf{k},\sigma }^{\dagger }$ and $a_{%
\mathbf{k},\sigma }$ are the creation and annihilation operators; spin index
$\sigma $ is introduced to normalize the chemical potential for total
numbers of electrons. In our case
\begin{equation}
\sum\limits_{\mathbf{k},\sigma }\left\langle a_{\mathbf{k},\sigma }^{\dagger
}a_{\mathbf{k},\sigma }\right\rangle \equiv \sum\limits_{\mathbf{k},\sigma
}f_{\mathbf{k},\sigma }=1.  \label{eq21}
\end{equation}

Now we can perform the renormalization procedure (\ref{eq5a}) with operator (%
\ref{eq19}) and represent (second-quantized) interaction $V$ (\ref{eq9}) in
the form
\begin{eqnarray}
&&\Omega _{X,Y} =-\frac \alpha {S_0}\sum _{\stackrel {\mathbf{k},\mathbf{k}%
^{\prime },\sigma }{\mathbf{k}\neq
\mathbf{k'}}}\sum\limits_jS_{X,Y}^j\left| \psi _{\perp }(\mathbf{R}%
_j)\right| ^2 \times \nonumber \\
&&\times e^{i\left( \mathbf{k}-\mathbf{k}^{\prime }\right) \mathbf{R}%
_j}a_{\mathbf{k},\sigma }^{\dagger }a_{\mathbf{k}^{\prime },\sigma },
\label{eq22} \\
&&\Omega _Z =-\frac \alpha {S_0}%
{\sum _{\stackrel {\mathbf{k},\mathbf{k'},\sigma }{\mathbf{k}\neq
\mathbf{k'}}} \sum\limits_jS_Z^j\left| \psi _{\perp
}(\mathbf{R}_j)\right| ^2e^{i\left( \mathbf{k}-\mathbf{k}^{\prime
}\right) \mathbf{R}_j}a_{\mathbf{k},\sigma }^{\dagger
}a_{\mathbf{k}^{\prime },\sigma }}.  \nonumber
\end{eqnarray}
Here $S_0$ is an area of a sample, $\psi _{\perp }(\mathbf{R}_j)$ is a
perpendicular (to a plane of a structure) component of confinement wave
function. The term with $\mathbf{k}\neq \mathbf{k}^{\prime }$ is excluded
from Eq. (\ref{eq22}) due renormalization procedure (\ref{eq5a}). This
procedure yields following Hamiltonian of dynamic subsystem:
\begin{equation}
H_S\equiv \omega s_{ez}=(\omega _e+G_{eZ})s_{ez}.  \label{eq23}
\end{equation}
Eq.(\ref{eq23}) defines spin splitting $\omega $ in a total field.
This field consists of external magnetic field (with Zeeman frequency $%
\omega _e$ ) and exchange molecular field
\begin{eqnarray}
&&\mathbf{G}_e=-\frac \alpha {S_0}\sum\limits_j\left\langle \mathbf{S}%
^j\right\rangle \left| \psi (z_j^{\prime })\right| ^2=\nonumber \\
&&=-\alpha \left\langle
\mathbf{S}\right\rangle \int\limits_{-\infty }^\infty n_m(z^{\prime })\left|
\psi (z^{\prime })\right| ^2dz^{\prime }.  \label{eq24}
\end{eqnarray}
Here $n_m(z^{\prime })$ is a local concentration of magnetic ions that is
assumed to be the function of coordinate $z$ directed along the growth axis $%
Z^{\prime }$. In the case of magnetic QW with width $L_w$ and non-magnetic
barriers, the integration should be performed between $-L_w/2$ and $L_w/2$,
while $n_m(z)$ should be replaced by average concentration $n_m$ in the QW.

We are looking for the Fourier transformation of Green's function
\[
\gamma _{\mu ,\nu }\left( \omega \right) =\left\langle \Omega _\mu \left(
t\right) ;\Omega _\nu \right\rangle _\omega ,
\]
which (according to Eq. (\ref{eq22})) equals to
\end{multicols}
\widetext
\noindent\rule{20.5pc}{0.1mm}\rule{0.1mm}{1.5mm}\hfill

\[
\gamma _{\mu ,\nu }\left( \omega \right) =\frac{\alpha ^2}{S_0^2}%
\sum\limits_{\mathbf{k},\mathbf{k}^{\prime },\sigma }\sum\limits_{\mathbf{k}%
_1,\mathbf{k}_1^{\prime },\sigma _1}\sum\limits_{j,j^{\prime
}}\left| \psi (z_j)\right| ^2\left| \psi (z_{j^{\prime }})\right|
^2e^{i\left( \mathbf{k}-\mathbf{k}^{\prime }\right) \mathbf{R}_j+i\left(
\mathbf{k}_1-\mathbf{k}_1^{\prime }\right) \mathbf{R}_{j^{\prime }}}K_{%
\mathbf{k},\sigma ;\mathbf{k}^{\prime },\sigma ;\mathbf{k}_1,\sigma_1 ;\mathbf{%
k}_1^{\prime },\sigma_1 }^{\mu ,j;\nu ,j^{\prime }}\left( \omega \right) ,
\]
\hfill\rule[-1.5mm]{0.1mm}{1.5mm}\rule{20.5pc}{0.1mm}
\begin{multicols}{2}
\narrowtext
where
\begin{eqnarray}
&&K_{\mathbf{k},\sigma ;\mathbf{k}^{\prime },\sigma ;\mathbf{k}_1,\sigma_1 ;%
\mathbf{k}_1^{\prime },\sigma_1 }^{\mu ,j;\nu ,j^{\prime }}\left( \omega
\right) = \nonumber \\
&&=\left\langle S_\mu ^ja_{\mathbf{k},\sigma }^{\dagger }a_{\mathbf{k}%
^{\prime },\sigma }\left( t\right) ;S_\nu ^{j^{\prime }}a_{\mathbf{k}%
_1,\sigma _1}^{\dagger }a_{\mathbf{k}_1^{\prime },\sigma _1}\right\rangle
_\omega .  \label{eq25}
\end{eqnarray}
The explicit form for equation of motion (\ref{A4}) for the
correspondent Green's function $\omega G_{q}^{\mu,\nu}$
($\mu,\nu=X,Y,Z$; $q$ is a rest of indexes that determines the
correlation function (\ref{eq25})) is approximated by the
following system of equations
\end{multicols}
\widetext
\noindent\rule{20.5pc}{0.1mm}\rule{0.1mm}{1.5mm}\hfill
\begin{eqnarray}
&&\omega G_{q}^{X,X}\left( \omega \right)  =\frac{\delta
_{j,j^{\prime }}\delta _{\mathbf{k},\mathbf{k}_1^{\prime }}\delta _{\mathbf{k%
}^{\prime },\mathbf{k}_1}\delta _{\sigma ,\sigma ^{\prime }}}{2\pi }%
\left\langle S_X^2\right\rangle \left[ f_{\mathbf{k},\sigma }(1-f_{\mathbf{k}%
^{\prime },\sigma })+f_{\mathbf{k}^{\prime },\sigma }(1-f_{\mathbf{k},\sigma
})\right] +  \label{eq26} \\
&&+\left( \varepsilon _{\mathbf{k}^{\prime }}-\varepsilon _{\mathbf{k}%
}\right) G_{q}^{X,X}\left( \omega \right) -i\omega
_0G_{q}^{Y,X}\left( \omega \right) ;  \nonumber \\
&&\omega G_{q}^{Y,X}\left( \omega \right)  =i\frac{\delta
_{j,j^{\prime }}\delta _{\mathbf{k},\mathbf{k}_1^{\prime }}\delta _{\mathbf{k%
}^{\prime },\mathbf{k}_1}\delta _{\sigma ,\sigma ^{\prime }}}{2\pi }\frac{%
\left\langle S_Z\right\rangle }2\left[ f_{\mathbf{k}^{\prime },\sigma }-f_{%
\mathbf{k},\sigma }\right] +
\left( \varepsilon _{\mathbf{k}^{\prime }}-\varepsilon _{\mathbf{k}%
}\right) G_{q}^{Y,X}\left( \omega \right) +i\omega
_0G_{q}^{X,X}\left( \omega \right) ;  \label{eq27}
\end{eqnarray}
\hfill\rule[-1.5mm]{0.1mm}{1.5mm}\rule{20.5pc}{0.1mm}
\begin{multicols}{2}
\narrowtext
where we use $\left\langle S_XS_Y\right\rangle =-\left\langle
S_YS_X\right\rangle =i\left\langle S_Z\right\rangle /2$; the occupation
numbers $f_{\mathbf{k},\sigma }$ are determined by Eq.(\ref{eq21}).
According to Eq.(\ref{A3}), the imaginary part of the $G_{q}^{\mu,\nu
}$ is important for calculation of $\mathbf{\Gamma }$. So,
taking into account that $\omega =\omega +i\epsilon $ with infinitesimal $%
\epsilon $, one can use following identity in the solution of Eqs (\ref{eq26}%
,\ref{eq27}):
\end{multicols}
\widetext
\noindent\rule{20.5pc}{0.1mm}\rule{0.1mm}{1.5mm}\hfill

\begin{eqnarray}
\frac{\omega -\left( \varepsilon ^{\prime }-\varepsilon \right) }{\left[
\omega -\left( \varepsilon ^{\prime }-\varepsilon \right) \right] ^2-\omega
_0^2} &\rightarrow &-i\frac \pi 2\left[ \delta \left( \omega -\left(
\varepsilon ^{\prime }-\varepsilon \right) -\omega _0\right) +\delta \left(
\omega -\left( \varepsilon ^{\prime }-\varepsilon \right) +\omega _0\right)
\right] ;  \nonumber  \label{eq28} \\
\frac{\omega _0}{\left[ \omega -\left( \varepsilon ^{\prime }-\varepsilon
\right) \right] ^2-\omega _0^2} &\rightarrow &-i\frac \pi 2\left[ \delta
\left( \omega -\left( \varepsilon ^{\prime }-\varepsilon \right) -\omega
_0\right) -\delta \left( \omega -\left( \varepsilon ^{\prime }-\varepsilon
\right) +\omega _0\right) \right] .  \label{eq28}
\end{eqnarray}

The final result for Fourier transformation of correlation function looks
like
\begin{eqnarray}
&&K_{\mathbf{k},\sigma ;\mathbf{k}^{\prime },\sigma ;\mathbf{k}_1,\sigma_1 ;%
\mathbf{k}_1^{\prime },\sigma_1 }^{X,j;X,j^{\prime }}\left( \omega \right)
=\frac {i}{2n}\left(G_{q }^{X,X}(\omega +i\epsilon )-G_{q }^{X,X}(\omega -i\epsilon )\right);\nonumber \\
&&K_{\mathbf{k},\sigma ;\mathbf{k}^{\prime },\sigma ;\mathbf{k}_1,\sigma_1 ;%
\mathbf{k}_1^{\prime },\sigma_1 }^{X,j;X,j^{\prime }}\left( \omega \right)  =
\frac{\delta _{j,j^{\prime }}\delta _{\mathbf{k},\mathbf{k}_1^{\prime
}}\delta _{\mathbf{k}^{\prime },\mathbf{k}_1}\delta _{\sigma ,\sigma_1
}}{4n}\Biggl\{\left\langle S_X^2\right\rangle [f_{\mathbf{k},\sigma
}(1-f_{\mathbf{k}^{\prime },\sigma })+f_{\mathbf{k}^{\prime },\sigma }(1-f_{%
\mathbf{k},\sigma })]\times  \label{eq29} \\
&&\times \left[ \delta \left( \omega -\left( \varepsilon ^{\prime }-\varepsilon
\right) -\omega _0\right) +\delta \left( \omega -\left( \varepsilon ^{\prime
}-\varepsilon \right) +\omega _0\right) \right] +  \nonumber \\
&&+\frac{\left\langle S_Z\right\rangle }{2n}\left[ f_{\mathbf{k}^{\prime
},\sigma }-f_{\mathbf{k},\sigma }\right] \left[ \delta \left( \omega -\left(
\varepsilon ^{\prime }-\varepsilon \right) -\omega _0\right) -\delta \left(
\omega -\left( \varepsilon ^{\prime }-\varepsilon \right) +\omega _0\right)
\right] \Biggr\};  \nonumber \\
&&K_{\mathbf{k},\sigma ;\mathbf{k}^{\prime },\sigma ;\mathbf{k}_1,\sigma_1 ;%
\mathbf{k}_1^{\prime },\sigma_1 }^{Y,j;X,j^{\prime }}\left( \omega \right)  =
\frac{i}{4n}\delta _{j,j^{\prime }}\delta _{\mathbf{k},\mathbf{k}_1^{\prime
}}\delta _{\mathbf{k}^{\prime },\mathbf{k}_1}\delta _{\sigma ,\sigma_1
}\Biggl\{\frac{\left\langle S_Z\right\rangle }2\left[ f_{\mathbf{k}%
^{\prime },\sigma }-f_{\mathbf{k},\sigma }\right]\times   \nonumber \\
&&\times \left[ \delta \left( \omega -\left( \varepsilon ^{\prime }-\varepsilon
\right) -\omega _0\right) +\delta \left( \omega -\left( \varepsilon ^{\prime
}-\varepsilon \right) +\omega _0\right) \right] +  \label{eq30} \\
&&+\frac{\left\langle S_X^2\right\rangle }n[f_{\mathbf{k},\sigma }(1-f_{%
\mathbf{k}^{\prime },\sigma })+f_{\mathbf{k}^{\prime },\sigma }(1-f_{\mathbf{%
k},\sigma })]\left[ \delta \left( \omega -\left( \varepsilon ^{\prime }-\varepsilon
\right) -\omega _0\right) -\delta \left( \omega -\left( \varepsilon ^{\prime
}-\varepsilon \right) +\omega _0\right) \right] \}.  \nonumber
\end{eqnarray}
with $\varepsilon  =\varepsilon _{\mathbf{k}},\varepsilon ^{\prime
}=\varepsilon _{\mathbf{k}^{\prime }}$.
\hfill\rule[-1.5mm]{0.1mm}{1.5mm}\rule{20.5pc}{0.1mm}
\begin{multicols}{2}
\narrowtext

We assume that neither 2D structure stress, nor internal magnetic
field affects the transversal magnetization (which means
$\left\langle
S_X^2\right\rangle =\left\langle S_Y^2\right\rangle $). Thus $K_{\mathbf{k}%
,\sigma ;\mathbf{k}^{\prime },\sigma ;\mathbf{k}_1,\sigma_1 ;\mathbf{k}%
_1^{\prime },\sigma_1 }^{Y,j;Y,j^{\prime }}\left( \omega \right) =K_{\mathbf{k}%
,\sigma ;\mathbf{k}^{\prime },\sigma ;\mathbf{k}_1,\sigma_1 ;\mathbf{k}%
_1^{\prime },\sigma_1 }^{X,j;X,j^{\prime }}\left( \omega \right) $ and $K_{%
\mathbf{k},\sigma ;\mathbf{k}^{\prime },\sigma ;\mathbf{k}_1,\sigma_1 ;\mathbf{%
k}_1^{\prime },\sigma_1 }^{X,j;Y,j^{\prime }}\left( \omega \right) =-K_{%
\mathbf{k},\sigma ;\mathbf{k}^{\prime },\sigma ;\mathbf{k}_1,\sigma_1 ;\mathbf{%
k}_1^{\prime },\sigma_1 }^{Y,j;X,j^{\prime }}\left( \omega \right) $. Latter
equality shows that the exchange scattering corresponds just to the case of
relaxation matrix (\ref{eq14a}). Direct calculation of off-diagonal matrix
elements shows that $K_{\mathbf{k},\sigma ;\mathbf{k}^{\prime },\sigma ;%
\mathbf{k}_1,\sigma_1 ;\mathbf{k}_1^{\prime },\sigma_1 }^{X,j;Z,j^{\prime
}}\left( \omega \right) =K_{\mathbf{k},\sigma ;\mathbf{k}^{\prime },\sigma ;%
\mathbf{k}_1,\sigma_1 ;\mathbf{k}_1^{\prime },\sigma_1 }^{Y,j;Z,j^{\prime
}}\left( \omega \right) =0$ since $\left\langle S_X\right\rangle
=\left\langle S_Y\right\rangle =0.$ In a similar manner, one can find
\begin{eqnarray}
&&K_{\mathbf{k},\sigma ;\mathbf{k}^{\prime },\sigma ;\mathbf{k}_1,\sigma_1 ;%
\mathbf{k}_1^{\prime },\sigma_1 }^{Z,j;Z,j^{\prime }}\left( 0\right) =\frac 12{%
\delta _{j,j^{\prime }}\delta _{\mathbf{k},\mathbf{k}_1^{\prime }}\delta _{%
\mathbf{k}^{\prime },\mathbf{k}_1}\delta _{\sigma ,\sigma_1}}%
\left\langle S_Z^2\right\rangle \times \nonumber \\
&&\times [f_{\mathbf{k},\sigma }(1-f_{\mathbf{k}%
^{\prime },\sigma })+f_{\mathbf{k}^{\prime },\sigma }(1-f_{\mathbf{k},\sigma
})]\delta \left( \varepsilon ^{\prime }-\varepsilon \right) .  \label{eq31}
\end{eqnarray}

The relaxation parameters in the matrix (\ref{eq14a}) can now be obtained
with the help of (\ref{eq29},\ref{eq30}). Eq.(\ref{eq21}) reads now
\begin{equation}
\sum\limits_\sigma f_{\mathbf{k},\sigma }=\frac{2\pi \hbar ^2}{S_0mT}%
e^{-\varepsilon _{\mathbf{k}}/T},  \label{eq33}
\end{equation}
where $m$ is the in-plane effective mass of the carriers. Substitution of
Eq.(\ref{eq31}) to the formula $\pi n\left\langle \Omega _Z\left( \tau
\right) \Omega _Z\right\rangle _0$ gives following result in the case of $f_{%
\mathbf{k},\sigma }\ll 1$ :
\begin{equation}
\pi \gamma _{zz}^0=\pi \frac{\alpha ^2}{S_0^2}\stackrel{(\mathbf{k}\neq
\mathbf{k}^{\prime })}{\sum\limits_{\mathbf{k},\mathbf{k}^{\prime },\sigma }}%
\sum\limits_j\left\langle S_Z^2\right\rangle \left| \psi (z_j)\right| ^4[f_{%
\mathbf{k},\sigma }+f_{\mathbf{k}^{\prime },\sigma }]\delta (\varepsilon
-\varepsilon ^{\prime }).  \label{eq32}
\end{equation}
Calculation of sums in Eq.(\ref{eq32}) yields
\begin{equation}
\frac \pi \hbar \gamma _{zz}^0=I\frac{\alpha ^2m}{2L_w\hbar ^3}%
n_m\left\langle S_Z^2\right\rangle ,  \label{eq34}
\end{equation}
where dimensionless parameter $I$ reflects the overlap of $\psi $-function
and magnetic ions:
\begin{equation}
I=L_w\int\limits_{-\infty }^\infty \frac{n_m(z^{\prime })}{n_m}\left| \psi
(z^{\prime })\right| ^4dz^{\prime }  \label{eq35}
\end{equation}
In the case of infinitely deep QW containing the magnetic ions with
concentration $n_m$, the integration in (\ref{eq35}) gives $I=3/2$.

To obtain other elements of the matrix (\ref{eq14a}) we should find $\gamma
_{xx}$ and $\gamma _{xy}$ according to Eqs(\ref{eq29}),(\ref{eq30}). It can
be shown that contribution of a second term in Eq.(\ref{eq29}) is less then
that of first one, while $\gamma _{xy}\ll \gamma _{xx}=\gamma _{yy}$, so
that we can find
\end{multicols}
\widetext
\noindent\rule{20.5pc}{0.1mm}\rule{0.1mm}{1.5mm}\hfill
\begin{equation}
\pi n\gamma _{xx} =\pi \frac{\alpha ^2}{S_0^2}\stackrel{(\mathbf{k}\neq
\mathbf{k}^{\prime })}{\sum\limits_{\mathbf{k},\mathbf{k}^{\prime },\sigma }}%
\sum\limits_j\frac{\left\langle S_X^2\right\rangle }2\left| \psi
(z_j)\right| ^4[f_{\mathbf{k},\sigma }+f_{\mathbf{k}^{\prime },\sigma
}][\delta (\omega -(\varepsilon -\varepsilon ^{\prime })-\omega _0)+
\delta (\omega -(\varepsilon -\varepsilon ^{\prime })+\omega _0)]. \label{eq36}
\end{equation}
\hfill\rule[-1.5mm]{0.1mm}{1.5mm}\rule{20.5pc}{0.1mm}
\begin{multicols}{2}
\narrowtext
Sums in Eq.(\ref{eq36}) are evaluated similar to those in Eq.(\ref{eq32}):
\begin{equation}
\frac \pi \hbar n\gamma _{xx}=I\frac{\alpha ^2m}{L_w\hbar ^3}n_m\frac{%
\left\langle S_X^2\right\rangle }2F\left( \frac \omega T,\frac{\omega _0}%
T\right) ,  \label{eq37}
\end{equation}
where we introduce the function
\begin{equation}
F(x,y) =\frac 14\left[ 2+\exp
\left( -|x+y| \right) +\exp \left( -|x-y|\right) \right]   \label{eq37e}
\end{equation}
that decreases monotonically from 1 to $1/2$ and reflects the suppression of
electron spin fluctuations by a magnetic field. Thus, the kinetic equations(%
\ref{eq14a}) transforms to the Bloch equations(\ref{eq2a}) with following
relaxation parameters
\begin{equation}
\frac 1{T_1}=I\frac{\alpha ^2m}{L_w\hbar ^3}n_m\left\langle
S_X^2\right\rangle F\left( \frac \omega T,\frac{\omega _0}T\right) ,
\label{eq37a}
\end{equation}
\begin{equation}
\frac 1{T_2}=I\frac{\alpha ^2m}{L_w\hbar ^3}n_m\left\{ \frac{\left\langle
S_Z^2\right\rangle }2+\frac{\left\langle S_X^2\right\rangle }2F\left( \frac
\omega T,\frac{\omega _0}T\right) \right\} .  \label{eq37b}
\end{equation}
The right hand parts of Eqs (\ref{eq34}) and (\ref{eq37}) equal to those of Eqs (%
\ref{eq37a}) and (\ref{eq37b}) respectively at zero magnetic field
as it should be expected.
Moreover, as $\omega \rightarrow 0,\omega _0\rightarrow 0$, Eqs (\ref{eq37a}%
),(\ref{eq37b}) transform just to the double rate $W$ of spin flip
relaxation obtained with Golden Fermi
Rule.\cite{Bastard1},\cite{SK}

With magnetic field increasing, one can obtain in the limit $\omega ,\omega
_0\gg T,$
\begin{equation}
\frac{T_1}{T_2}=2S+\frac 12.  \label{eq38}
\end{equation}
Eq.(\ref{eq38}) shows that rate of phase relaxation exceeds the longitudinal
relaxation rate with factor about one order of magnitude in the case of ions
$Mn^{2+}$ ($S=5/2$). In the case of equidistant Zeeman splitting $\omega _0$
of the magnetic ions, the magnetic field dependence of spin square averages
is:
\begin{eqnarray}
\left\langle S_X^2\right\rangle &=&\frac 12\coth \frac{\omega _0}{2T}\text{b}%
_S(\frac{\omega _0}T),  \label{eq39} \\
\left\langle S_Z^2\right\rangle &=&S(S+1)-2\left\langle S_X^2\right\rangle ,
\nonumber
\end{eqnarray}
where b$_S(x)=-\left\langle S_Z\right\rangle =SB_S(Sx)$ is non-normalized
Brillouin function:
\begin{equation}
\text{b}_S(x)=(S+\frac 12)\coth (S+\frac 12)x-\frac 12\coth \frac x2
\label{eq39a}
\end{equation}
Figure 1 shows the magnetic field dependence of the longitudinal and
transversal relaxation times calculated with the help of Eqs (\ref{eq37a}), (%
\ref{eq37b}), (\ref{eq39}), (\ref{eq39a}). One can see that
magnetic field suppresses the longitudinal relaxation while it
enhances transversal relaxation for $S>1/2$. In the limit of
saturated magnetic field, $\omega
,\omega _0>>T$, the decreasing of $T_2$ reaches the factor $\frac 23\frac{S+1%
}{S+1/2}$. Note, that the decreasing of transversal relaxation
time in a magnetic field was observed for $Mn$ spins but was not
explained till now.

At the end of this section we estimate the relaxation time of the electron
in $Cd_{1-x}Mn_xTe$ QW with x=0.017 and $L_w=8$ nm. At zero magnetic field
one can find $\tau _e=5$ ps that is quite close to experimental value
observed recently via time resolved magneto-optic Kerr effect.\cite{Denis1}

\section{Dyakonov-Perel mechanism}

In a number of works \cite{DP},\cite{DKa},\cite{Pikus}, it have been shown
that a fluctuating effective magnetic field accompaning the electron
momentum scattering in the semiconductors with non-centrosymmetric lattice
results in rather efficient spin relaxation. In this section we show how
Dyakonov-Perel mechanism can be incorporated into formalism developed above.

Hamiltonian of dissipative subsystem is assumed to have the form

\begin{equation}
H_L=\sum\limits_{\mathbf{k},\sigma }\varepsilon _{\mathbf{k}}a_{\mathbf{k}%
,\sigma }^{\dagger }a_{\mathbf{k},\sigma }+\sum\limits_{\mathbf{k},\mathbf{k}%
^{\prime },\sigma }V_{\mathbf{k},\mathbf{k}^{\prime }}a_{\mathbf{k},\sigma
}^{\dagger }a_{\mathbf{k}^{\prime },\sigma }.  \label{eq40}
\end{equation}
Second term in Eq.(\ref{eq40}) has symbolic meaning: we assume $V_{\mathbf{k}%
,\mathbf{k}^{\prime }}$ to be an operator responsible for electron
scattering by the impurities, phonons, or electron-electron collisions. We
should also introduce the electron scattering changing randomly an effective
magnetic field $\mathbf{B}=\mathbf{b}(\mathbf{k})/g_e\mu $ that corresponds
to band spin splitting
\begin{equation}
H_{int}=\sum\limits_{\mathbf{k},\sigma }\mathbf{s}_e\mathbf{b}(\mathbf{k})a_{%
\mathbf{k},\sigma }^{\dagger }a_{\mathbf{k},\sigma }.  \label{eq41}
\end{equation}
Therefore, the operator of this field reads
\begin{equation}
\Omega _\mu =\sum\limits_{\mathbf{k},\sigma }b_\mu (\mathbf{k})a_{\mathbf{k}%
,\sigma }^{\dagger }a_{\mathbf{k},\sigma }.  \label{eq42}
\end{equation}
The specific form of the function $b_\mu (\mathbf{k})$ is determined by a
band Hamiltonian (see, for instance, Ref.\cite{Pikus}) and depends on
dimensionality of the system under consideration (Ref.\cite{DKa}).

We introduce the Green's function

\begin{equation}
\tilde G\equiv \tilde G_{2,3}^{}=\left\langle \left\langle a_{\mathbf{k}%
_2}^{\dagger }a_{\mathbf{k}_3};a_{\mathbf{k}_1}^{\dagger }a_{\mathbf{k}%
_1}\right\rangle \right\rangle _\omega .  \label{eq43}
\end{equation}
For simplicity, here we suppress spin indices while $1\equiv \mathbf{k}%
_{1,2}\equiv \mathbf{k}_2$, etc. Equation (\ref{A4}) with Hamiltonian (\ref
{eq40}) assumes the form
\begin{eqnarray}
&&\left[ \omega -\left( \varepsilon _1-\varepsilon _2\right) \right] \tilde
G_{2,3}=\frac{\delta _{2,1}\delta _{3,1}}\pi f_1(1-f_1)+\nonumber \\
&&+ \sum_4\left(
V_{3,4}\tilde G_{2,4}-V_{4,2}\tilde G_{4,3}\right) .  \label{eq44}
\end{eqnarray}
We are looking for $\tilde G_{1,1}$ that can be represented via $\tilde
G_{2,3}$. Solution of the Eq.\ref{eq44} depends strongly on $\mathbf{k,k}%
^{\prime }$ - dependences of operators $V_{\mathbf{k},\mathbf{k}^{\prime }}$%
. We assume that following general relation for Green's function holds (Ref.%
\cite{Zub})
\begin{equation}
\tilde G_{1,1}=\frac{f_1(1-f_1)}{\pi \left( \omega +i\Gamma _{1,1}\right) },
\label{eq45}
\end{equation}
where $\Gamma _{1,1}$ means the rate of electron moment relaxation due to
scattering processes described by second term in the Eq.(\ref{eq40}). Thus,
the relaxation parameters are defined by
\begin{equation}
\frac \pi \hbar n\gamma _{\mu ,\nu }=\sum\limits_{\mathbf{k}}b_\mu (\mathbf{k%
})b_\nu (\mathbf{k})\frac{\Gamma _{\mathbf{k},\mathbf{k}}/\hbar
}{\hbar ^2\omega ^2+\Gamma _{\mathbf{k},\mathbf{k}}^2}.
\label{eq46v}
\end{equation}

Let us consider the electron spin splitting caused by $k^3$ - terms in band
Hamiltonian (Ref.\cite{DP}). This mechanism had been considered for
2D-electron gas in the work\cite{DKa}. If external magnetic field is
directed along the axis of structure growth, one can obtain

\begin{equation}
b_\mu (\mathbf{k})=r_\alpha \kappa _\mu  \label{eq42v}
\end{equation}
with
\begin{equation}
r_\alpha =\frac{\alpha \hbar ^3}{m^{3/2}(2E_g)^{1/2}},
\label{eq43v}
\end{equation}
\begin{equation}
\kappa _x=-k_xq^2;\kappa _y=k_yq^2;\kappa _z=0,  \label{eq44a}
\end{equation}
where $m$ is in-plane effective mass, $q^2$ is defined via electron $\psi $-
function confined in a QW,
\begin{equation}
q^2=\left\langle \psi \left| -\frac{\partial ^2}{\partial
z^2}\right| \psi \right\rangle .  \label{eq45v}
\end{equation}

We see that only two components of relaxation parameters, $\gamma _{x,x}$
and $\gamma _{y,y}$ have nonzero values. The calculations with the help of
Eq. (\ref{eq33}) give rise to following expression for transversal
relaxation time
\begin{equation}
\frac 1{T_2}=\frac \pi \hbar n\gamma _{x,x}=\frac \pi \hbar n\gamma _{y,y}=%
\frac{\alpha ^2\hbar ^2q^4T}{2m^2E_g}\frac{\tau _e}{1+(\omega \tau _e)^2},
\label{eq46}
\end{equation}
where the electron scattering time is defined via averaged value of $\Gamma
_{\mathbf{k},\mathbf{k}}:$
\begin{equation}
\tau _e^{-1}=\hbar ^{-1}\left\langle \Gamma _{\mathbf{k},\mathbf{k}%
}\right\rangle .  \label{eq47}
\end{equation}

Note that approximation (\ref{eq47}) is not nessessary for calculations of $%
T_1$ and $T_2$. The dependence $\tau _e(\mathbf{k})$ can be taken into
account for more accurate theories (see Ref.\cite{Pikus}).

According to Eq.(\ref{eq14}), the longitudinal relaxation rate is
\begin{equation}
\frac 1{T_1}=\frac \pi \hbar n(\gamma _{x,x}+\gamma _{y,y})=\frac 2{T_2}.
\label{eq48}
\end{equation}
We can see that $T_1$can be shorter than $T_2$ in the systems with strong
anisotropy .

Other important case corresponds to magnetic field directed along $OZ$ axis
in QW plane while $OX$ axis is a perpendicular to this plane. One can find

\begin{equation}
\kappa _x=0;\kappa _y=-k_yq^2;\kappa _z=k_zq^2.  \label{eq49}
\end{equation}

In this case, three parameters describe electron spin relaxation:
\begin{eqnarray}
\frac 1{T_{2X}} &=&\frac{\alpha ^2\hbar ^2q^4T}{2m^2E_g}\tau _e\left(
1+\frac 1{1+(\omega \tau _e)^2}\right) ,  \nonumber \\
\frac 1{T_{2Y}} &=&\frac{\alpha ^2\hbar ^2q^4T}{2m^2E_g}\tau _e,  \nonumber
\\
\frac 1{T_1} &=&\frac{\alpha ^2\hbar ^2q^4T}{2m^2E_g}\frac{\tau _e}{%
1+(\omega \tau _e)^2}.  \label{eq50}
\end{eqnarray}
We can see that transversal relaxation (Eq.\ref{eq46}) can be
suppressed significantly by the strong enough magnetic field,
$\omega \tau _e>>1,$ due to dynamical averaging effect in the case
of Faraday geometry of experiment \cite{Ivch}. In the case of
Voigt configuration (Eqs (\ref{eq50})), magnetic field can
suppress the transversal relaxation time by no more than two
times. The effect of drastic magnetic anisotropy of spin
relaxation in QWs can be used to recognize the specific mechanism
under consideration. Actually, the typical scattering time of an
electron in semiconductors is about $\tau _e\sim 10^{-12}$. To
make the magnetic field anisotropy effect visible, one should
apply quite strong magnetic field to non-magnetic structure. On
the other hand, considered anisotropy effect can be amplified in
DMS structures giving rise to inequality $\omega >>\omega _0$ at
moderate magnetic fields (Eqs (\ref{eq23}), (\ref{eq24})).

\section{Conclusions}

In this work, we develop the microscopic theory of electron spin evolution
in semiconductors in terms of quantum kinetic equations for arbitrary
mechanism of spin relaxation. The relaxation term in these equations has
been derived as a matrix of correlation functions providing relatively short
correlation times ( as compared to spin relaxation times, $\tau <<T$) of
dissipative subsystem. Our theory permits to distinguish the phase and
energy relaxation processes and to capture the case of zero and small
magnetic fields.

\begin{figure}[th]
\vspace*{-5mm}
\centerline{\centerline{\psfig{figure=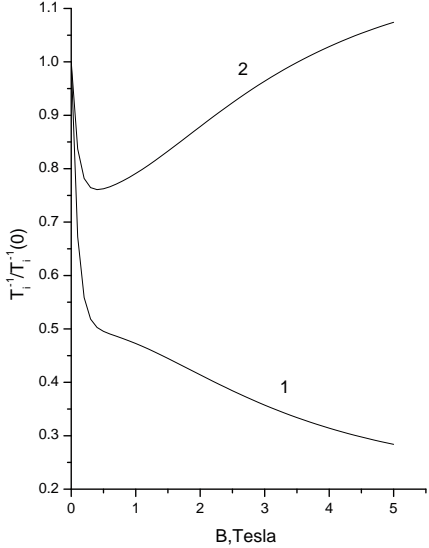,width=0.8\columnwidth}}}
\vspace*{0.5mm} \caption{Magnetic field dependencies of the rates
of longitudinal ($i=1$) and transversal ($i=2$) electron spin
relaxation at $T=4K$ in $Cd_{1-x}Mn_xTe$ ($x=0.017$) crystal.}
\end{figure}

We show that spin relaxation process due to exchange scattering in a
semimagnetic QW determines the longitudinal and transversal spin relaxation
times. Specific calculations of relaxation parameters show that only
longitudinal relaxation can be described by flip-flop processes which is
suppressed in a magnetic field. On the other hand, transversal relaxation is
due to the effective exchange field fluctuations, which increase with
magnetic field. Qualitatively, the latter effect has the same reason as
growth of line width of spin flip Raman scattering by shallow donors in
diluted magnetic semiconductors \cite{SemRy}.

The theory has been successfully applied to Dyakonov-Perel mechanism of
carrier spin relaxation in a QW. The significant difference in transversal
relaxation rates for parallel and perpendicular (to growth axis) magnetic
field orientations has been found.

Proposed approach to relaxation can be extended to any two-level system
described by fictitious spin 1/2. From this standpoint, the analysis of
heavy hole spin relaxation in QW in terms of Eqs (\ref{eq13},\ref{eq14},\ref
{eq14a}) promises new insight into this problem. The investigations of a
hole relaxation in QWs or uniaxial semiconductors will be reported elsewhere.

\section{Acknowledgement}
Author would like to thanks D. Scalbert for fruitful discussions. This work was partially supported by the Grant No. 09244106 of the Ministry
of Education and Science.

\appendix

\section{}

Equation (\ref{eq13}) reduces the problem of spin relaxation to calculations
of correlation functions of operators $\mathbf{\Omega }$ related to a heat
bath at a thermal equilibrium. Theory of correlation functions for
quantum-mechanical operators is well developed (see e.g. Ref.\cite{Zub}). In
this Appendix, we present few equations that are needed in subsequent
analysis of relaxation parameters.

Introduce first two-time temperature retarded Green's function (Ref.\cite{Zub}%
)
\begin{equation}
G(t,t^{\prime })\equiv \left\langle \left\langle \Omega _\mu ;\Omega _\nu
\right\rangle \right\rangle =-i\theta (t-t^{\prime })\left\langle \left[
\Omega _\mu \left( t\right) ,\Omega _\nu (t)\right] \right\rangle .
\label{A1}
\end{equation}
The bracket $\left[ \Omega _\mu \left( t\right) ,\Omega _\nu (t)\right] $
denotes a commutator or anticommutator for operators $\Omega _\mu \left(
t\right) $ and $\Omega _\nu (t)$. Fourier transformation of Green's function (%
\ref{A1}) reads
\begin{equation}
G(\omega )\equiv \left\langle \left\langle \Omega _\mu ;\Omega _\nu
\right\rangle \right\rangle _\omega =\frac 1{2\pi }\int\limits_{-\infty
}^\infty G(t,0)\exp (i\omega t)dt.  \label{A2}
\end{equation}
Fourier transformation of correlation functions (Eq.(\ref{eq11a}))
can be expressed in terms of that for Green's functions
Eq.(\ref{A2}):
\begin{equation}
\gamma _{\mu \nu }(\omega )=i{\lim _{\epsilon \to 0}}\frac{%
G(\omega +i\epsilon )-G(\omega -i\epsilon )}{2n(\omega )}.  \label{A3}
\end{equation}
So, the problem is reduced to explicit calculation of Green's functions in Eq.(%
\ref{A1}). The equation of motion for their Fourier component is \cite{Zub}:
\begin{equation}
\omega G(\omega )=\frac 1{2\pi }\left\langle \left[ \Omega _\mu ,\Omega _\nu
\right] \right\rangle +\left\langle \left\langle \left( \Omega _\mu
H_L-H_L\Omega _\mu \right) ;\Omega _\nu \right\rangle \right\rangle _\omega .
\label{A4}
\end{equation}
The solution of equation (\ref{A4}) permit to find the correlation function,
which, in turn, determines the spin relaxation according to Eqs (\ref{eq11a}%
), (\ref{eq14}) and (\ref{eq14a}).

\end{multicols}

\end{document}